\documentclass[aps,a4paper,amsmath,amssymb,twocolumn,
floatfix,superscriptaddress]{revtex4}
\usepackage{bm,amsmath,amssymb,amsfonts}
\usepackage[dvips]{graphicx}
\usepackage[utf8]{inputenc}
\usepackage[normalem]{ulem}
\usepackage{xcolor}
\definecolor{dblue}{rgb}{0.0,0.0,0.7}
\definecolor{dred}{rgb}{0.9,0.0,0.0}
\definecolor{dgreen}{rgb}{0,0.8,0.0}
\usepackage{hyperref}
\usepackage{cancel}
\hypersetup{
    pdfnewwindow=true,      
    colorlinks=true,       
    linkcolor=red,          
    citecolor=blue,        
    filecolor=blue,      
    urlcolor=blue        
}
\newcommand{\ie}{{\it i.e.}}
\newcommand{\eg}{{\it e.g.}}
\newcommand{\etal}{{\it et al.}}

\begin{document}
\title{Flux-induced strengthening of the magnetic couplings in a flat-band diamond chain}
\author{Biplab Pal}
\thanks{Corresponding author}
\email[E-mail: ]{biplab@nagalanduniversity.ac.in}
\affiliation{Department of Physics, School of Sciences, 
Nagaland University, Lumami-798627, Nagaland, India}
\author{Maxime Thumin}
\email[E-mail: ]{maxime.thumin@cea.fr}
\affiliation{Universit\'{e} Grenoble Alpes, CEA, IRIG-MEM-L SIM, 
F-38000 Grenoble, France}
\author{Georges Bouzerar}
\email[E-mail: ]{georges.bouzerar@neel.cnrs.fr}
\affiliation{Universit\'{e} Grenoble Alpes, CNRS, Institut NEEL, 
F-38042 Grenoble, France}
\date{\today}
\begin{abstract}
The physics in flat bands has emerged as an essential field in condensed matter physics where a plethora 
of phenomena can be unveiled, such as anomalous transport properties, superconductivity dominated by quantum 
geometry or exotic topological phases. Our goal here is to show that even in magnetic systems, the presence 
of flat bands can give rise to unexpected features. More precisely, we address the impact of an Aharonov-Bohm 
(AB) flux on the exchange couplings in magnetic diamond chains. The most remarkable result is the significant 
amplification of magnetic couplings at short distances induced by the AB flux, leading to a considerable 
increase in the thermal conductivity of the magnons. We have also shown that the flux-dependent decaying length 
of the couplings is connected to the quantum metric of the flat bands. Our results could be of interest for the 
control of magnetic properties in spintronic devices and relevant for the heat transport by magnons at the nanoscale 
in quantum technologies.
\end{abstract}
\maketitle
%

\section{Introduction}
\label{sec:Intro}
In recent years, flat band (FB) physics has become one of the most fascinating research topics in physics and in materials 
science~\cite{FB-Review-Flach,Review-Balents,FB-Review-Chen,FB-Review-Vicencio}. Due to the quenching of the kinetic energy, 
FBs provide an ideal platform to address a plethora of strongly correlated phenomena, such as ferromagnetism~\cite{FB-ferromagnetism-1,
FB-ferromagnetism-2,FB-ferromagnetism-3,mag-GB1}, superconductivity~\cite{FB-Peotta,FB-superconductivity-1, FB-superconductivity-2,
FB-superconductivity-3,FB-superconductivity-4}, or even the quantum Hall effect~\cite{Wen-PRL-2011,Das-Sarma-PRL-2011,Neupert-PRL-2011}. 
The area of physics in flat bands is not limited to an exciting playground for theoreticians; it is also experiencing significant 
development in experimental research. The interest for FBs does not cease to grow in various directions. 
FBs are attracting considerable attention in many research areas, including photonic lattices~\cite{Vicencio-PRL-2015,Mukherjee-PRL-2015,
Longhi-OL-2014,Xia-OL-2016}, Moir\'{e} superlattices~\cite{Review-Balents,LeRoy-Moire-FB-2,Li-Moire-FB-3}, topo-electrical 
circuits~\cite{Zhang-EC-FB-1,Zhang-EC-FB-2,Zhang-EC-FB-3,Flach-EC-FB} and quantum electrodynamics (QED) lattices as well~\cite{Circuit-QED-FB}.
Recently, FBs have also been proposed for the realization of quantum storage device~\cite{Rudolf-prl-2025}. From a technological 
application perspective, the search for new materials with FBs near the Fermi level is a crucial challenge for 
material science. Realistic studies of various materials using DFT calculations play an essential role in facilitating this 
quest~\cite{jiang-nanoscale19,zhang-prb19,jiang-natcomm19,Regnault-Nat-2022}. 

\begin{figure}[ht]
\centering
\includegraphics[clip, width=0.85\columnwidth]{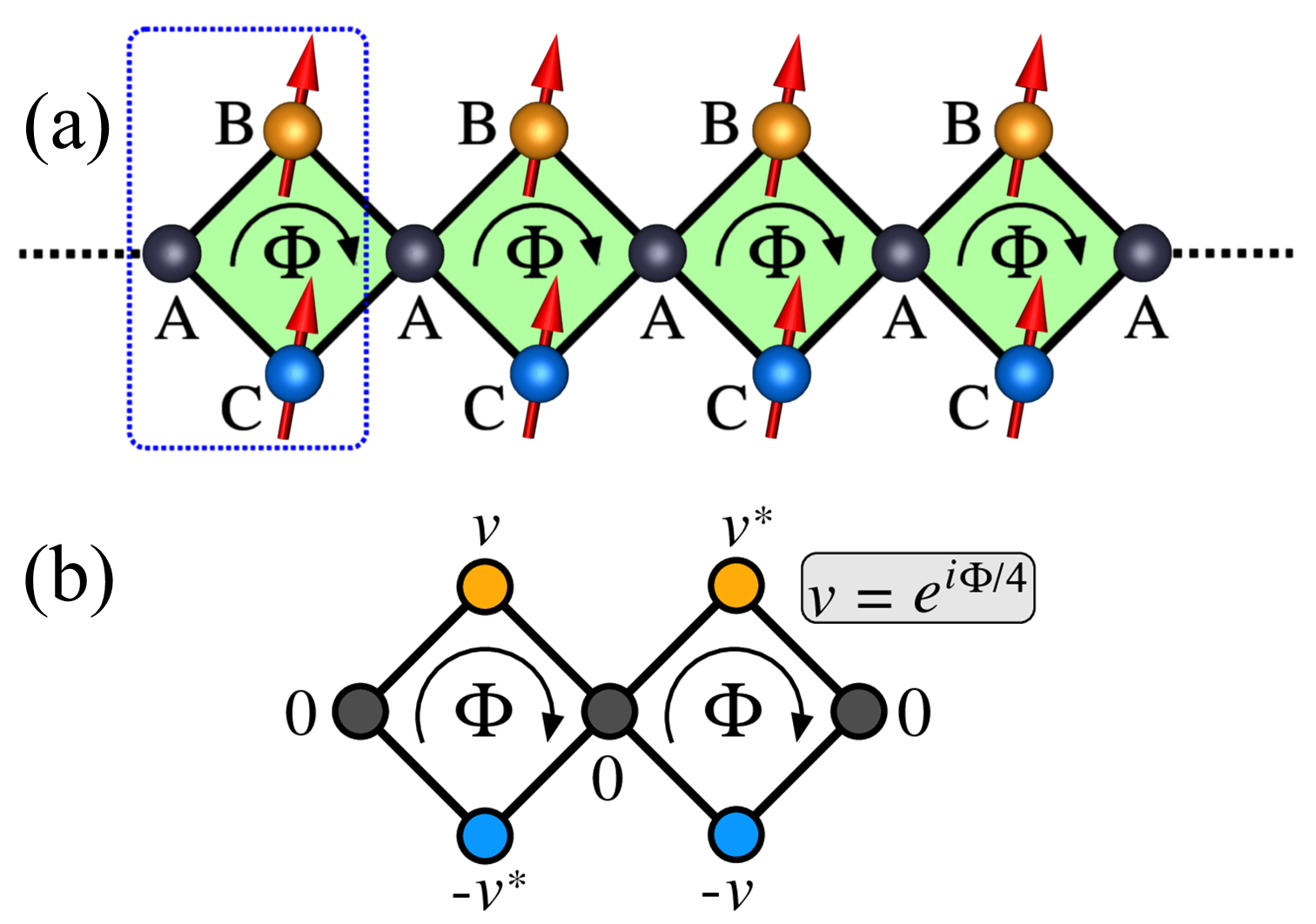}
\caption{(a) Representation of the magnetic diamond chain threaded by an 
external magnetic flux $\Phi$. The red arrows correspond to the localized 
classical spins located on $B$ and $C$ sites. The boxed area indicates 
the unit cell. The flux-dependent compact localized state (CLS) is 
depicted in panel (b). The CLS spreads over two unit cells, and its 
amplitudes in each site are shown.}
\label{lattice-n-CLS}
\end{figure}

In this study, our main objective is to reveal a fascinating physical phenomenon that occurs in magnetic systems with flat bands 
near the Fermi energy. We demonstrate that in such FB systems, it is possible to boost the magnetic couplings (at short distances) 
between localized spins by tuning an external magnetic flux. Furthermore, it will also be shown that the flux has a strong impact 
on the magnon branches and leads to a considerable increase in the thermal conductivity associated to the magnons. To illustrate 
this interesting phenomenon, we consider the case of the one-dimensional diamond lattice, also known as the rhombic lattice which 
consists of two different sublattices $\Lambda$ ($A$ sites) and $\Lambda'$ ($B$ and $C$ sites) as depicted in Fig.~\ref{lattice-n-CLS}(a). 
Because of the bipartiteness of the lattice, and since $|\Lambda'|-|\Lambda| = 1$, the diamond chain has a FB at $E=0$, whose 
eigenstates have a non-zero weight only on the orbitals of the sublattice $\Lambda'$ as shown in Fig.~\ref{lattice-n-CLS}(b). 
The classical localized spins are coupled locally to the orbitals of the sublattice $\Lambda'$.

Before we discuss the impact of the magnetic flux on the exchange couplings in the magnetic rhombic chains, we briefly recall 
that the effect of an Aharonov-Bohm (AB) flux~\cite{AB-Phase} in certain types of non-magnetic geometries has been widely 
studied in the context of single-particle localization, topological properties of flat bands, quantum transport, 
AB interferometer, etc.~\cite{Vidal-prl-2000,Vidal-prl-1998,Amrita-jpcm-2021,Biplab-prb-2018,Biplab-jpcm-2025,Biplab-pssrrl-2025,
Malay-PRB-2013,Malay-JPCM-2025,Biplab-PLA-2014,Tapan-Mishra-PRB-2025}. The reason 
is that the AB flux leads to an extreme localization phenomenon, known as the AB caging. This localization originates from 
destructive quantum interferences of different tunneling pathways. As a result, all dispersive bands of the single-particle 
spectrum become completely flat~\cite{Vidal-prl-2000,Vidal-prl-1998,Amrita-jpcm-2021}. The AB caging phenomenon has also been 
realized in experiments using the femtosecond laser-writing techniques in photonics~\cite{Sebabrata-prl-2018,Rodrigo-prl-2022,
Kremer-natcomm-2020}, as well as with ultracold atoms in optical lattices ~\cite{Longhi-prl-2022}. 

\section{Hamiltonian and spectrum}
\label{sec:Hamiltonian}
The Hamiltonian that describes electrons in interaction with the localized spins in the magnetic diamond chain threaded by a 
magnetic flux $\Phi$ reads,
\begin{equation} 
\widehat{H} =  \sum_{\left\langle i\lambda, j\lambda' \right\rangle, \sigma} \hspace{-0.2cm} 
\Big[ t^{\lambda\lambda'}_{ij}(\Phi) \hat{c}_{i\lambda\sigma}^{\dagger}\hat{c}_{j\lambda'\sigma} + \textrm{h.c.}\Big] 
 + J\sum_{i,\lambda \in \Lambda'}\hat{\bm s}_{i\lambda}\cdot {\bm S}_{i\lambda}.
\label{hamiltonian}
\end{equation}
$\hat{c}_{i\lambda\sigma}^{\dagger}$ creates an electron with spin $\sigma =\: \uparrow,\downarrow$ in the orbital $\lambda$ 
of the $i$-th cell. $\left\langle i\lambda,j\lambda' \right\rangle$ means nearest-neighbors. 
$t^{\lambda\lambda'}_{ij}(\Phi) = -te^{\pm i\Phi/4}$ depending on whether we go clockwise or anti-clockwise. 
$J$ is the local coupling between the localized classical spin ${\bm S}_{i\lambda} = S.{\bm e}_{i\lambda}$ at $\bm{r}_{i\lambda}$ 
(${\bm e}_{i\lambda}$ being a unit vector) and that of the itinerant carrier, 
$\hat{s}^{p}_{i\lambda} = \hat{c}_{i\lambda\alpha}^{\dagger} \left[ {\hat{\sigma}^{p}}\right]_{\alpha\beta} \hat{c}_{i\lambda\beta}$ 
where $p = x, y,\ \text{and}\ z$, and ${\hat{\sigma}^{p}}$ being the Pauli matrices. 
In what follows, we set $t = 1$ and $JS$ is expressed in units of $t$. Here, we focus essentially on the case of half-filled diamond 
We define `$a$' as the size of the unit cell (distance between the nearest-neighbor orbitals $A$), and for simplicity, 
we set $a = 1$. We remark that we consider classical spins only since in the case of ferromagnetism, the quantum versus 
classical nature of the spin is irrelevant. Indeed, in both cases, the true ground state is identical and quantum fluctuations 
do not play a critical role in contrast to antiferromagnetically coupled systems.

The lattice being bipartite and the localized spins being located on the same sublattice, at $T=0\:K$, the ground-state is 
ferromagnetic~\cite{Lieb,Saremi,mag-GB1}. Choosing the localized spin orientation along the $z$-axis, the second term in the 
Hamiltonian as given in Eq.~\eqref{hamiltonian} becomes $\frac{1}{2}z_{\sigma}JS \sum_{i,\lambda \in \Lambda'} 
\hat{c}_{i\lambda\sigma}^{\dagger}\hat{c}^{}_{i\lambda\sigma}$, where $z_{\sigma} = 1$ for $\sigma=\:\uparrow$ 
(resp. $z_{\sigma}=-1$ for $\sigma=\:\downarrow$). Here, we consider $J \ge 0$, the case $J \le 0 $ being equivalent. 
In momentum space, the Hamiltonian is diagonalized separately in the two spin sectors (see Appendix~\ref{AppendixA}). 
The eigenspectrum reads,
$E^{\sigma}_{\pm}=\frac{1}{2}\big[z_{\sigma}\frac{JS}{2} \pm \sqrt{\Delta_\Phi(k)}\big]$ and $E^{\sigma}_{0}=z_{\sigma}\frac{JS}{2}$, 
where $\Delta_{\Phi}(k)=\frac{1}{4}J^2S^2+16t^2[1+\cos(\frac{\Phi}{2})\cos(k)]$. 
The flat band survives even in the presence of the magnetic flux and its energy is flux independent. 
In the basis $(\hat{c}_{kA\sigma}^{\dagger},\: \hat{c}_{kB\sigma}^{\dagger},\: \hat{c}_{kC\sigma}^{\dagger})$, its eigenstate is 
$\vert \Psi^{\sigma}_{0}\rangle =\frac{1}{\sqrt{1+\cos(\frac{\Phi}{2})\cos(k)}}
\left(0,\: -\cos(\frac{k}{2}+\frac{\Phi}{4}),\: \cos(\frac{k}{2}-\frac{\Phi}{4})\right)$.
We remark that, real space FB eigenstates that possess a minimal finite support (non-vanishing weight on a restricted number of sites), 
known as the compact localized eigenstates (CLS) can be  constructed as well. In Fig.~\ref{lattice-n-CLS}(b), 
the CLS which depends on $\Phi$ only is depicted; its expression is $\vert CLS \rangle_{i,i+1} = \frac{1}{2} {\left[ v \vert B_i\rangle - 
v^* \vert C_i\rangle + v^* \vert B_{i+1}\rangle -v \vert C_{i+1}\rangle \right]}$, where $v = e^{i\Phi/4}$. We remark that, 
for a non-zero flux, the CLS spread is over two unit cells, while for $\Phi = 0$, it reduces to a single one.
\begin{figure}[ht]
\centering
\includegraphics[clip, width=0.49\columnwidth]{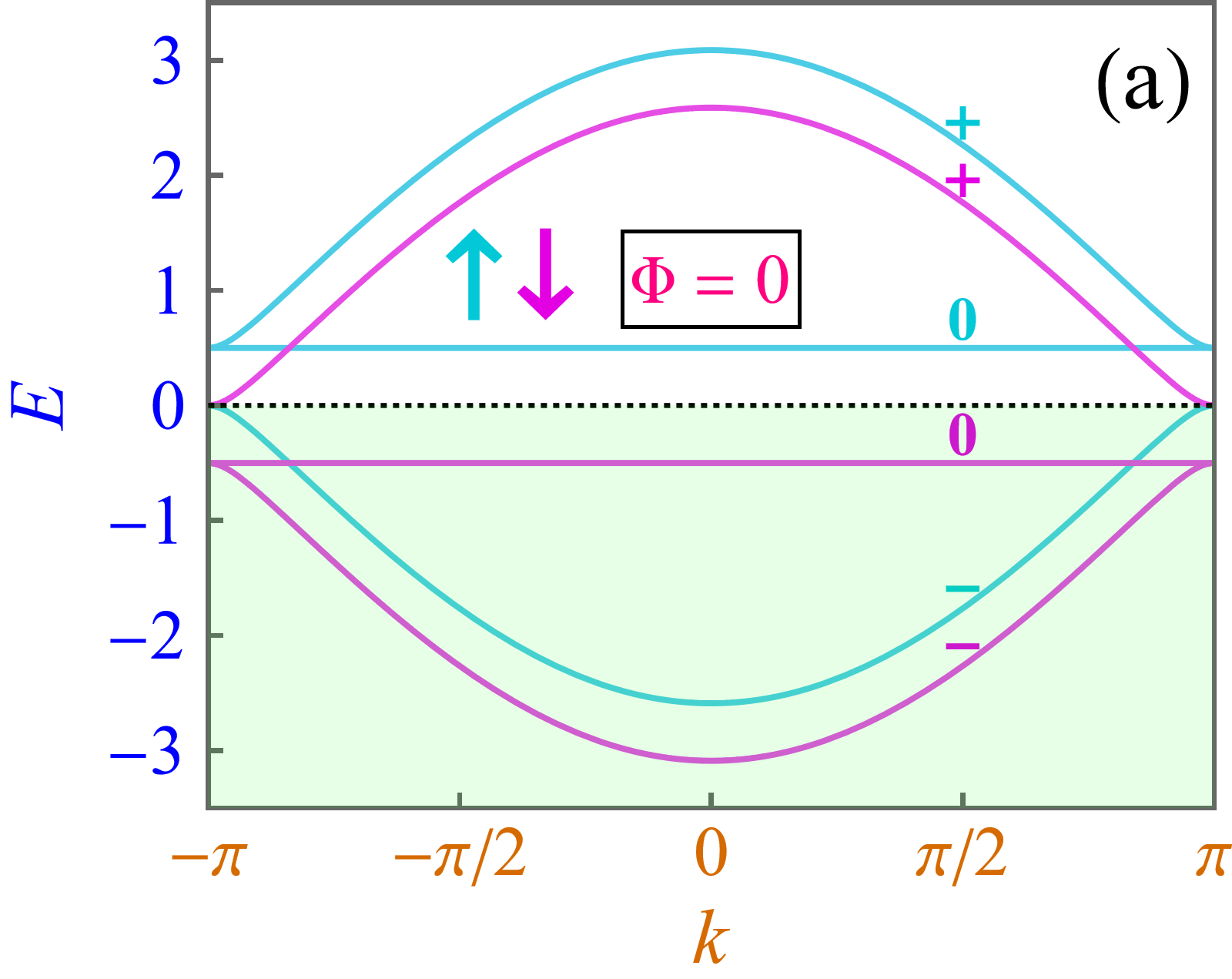}
\includegraphics[clip, width=0.49\columnwidth]{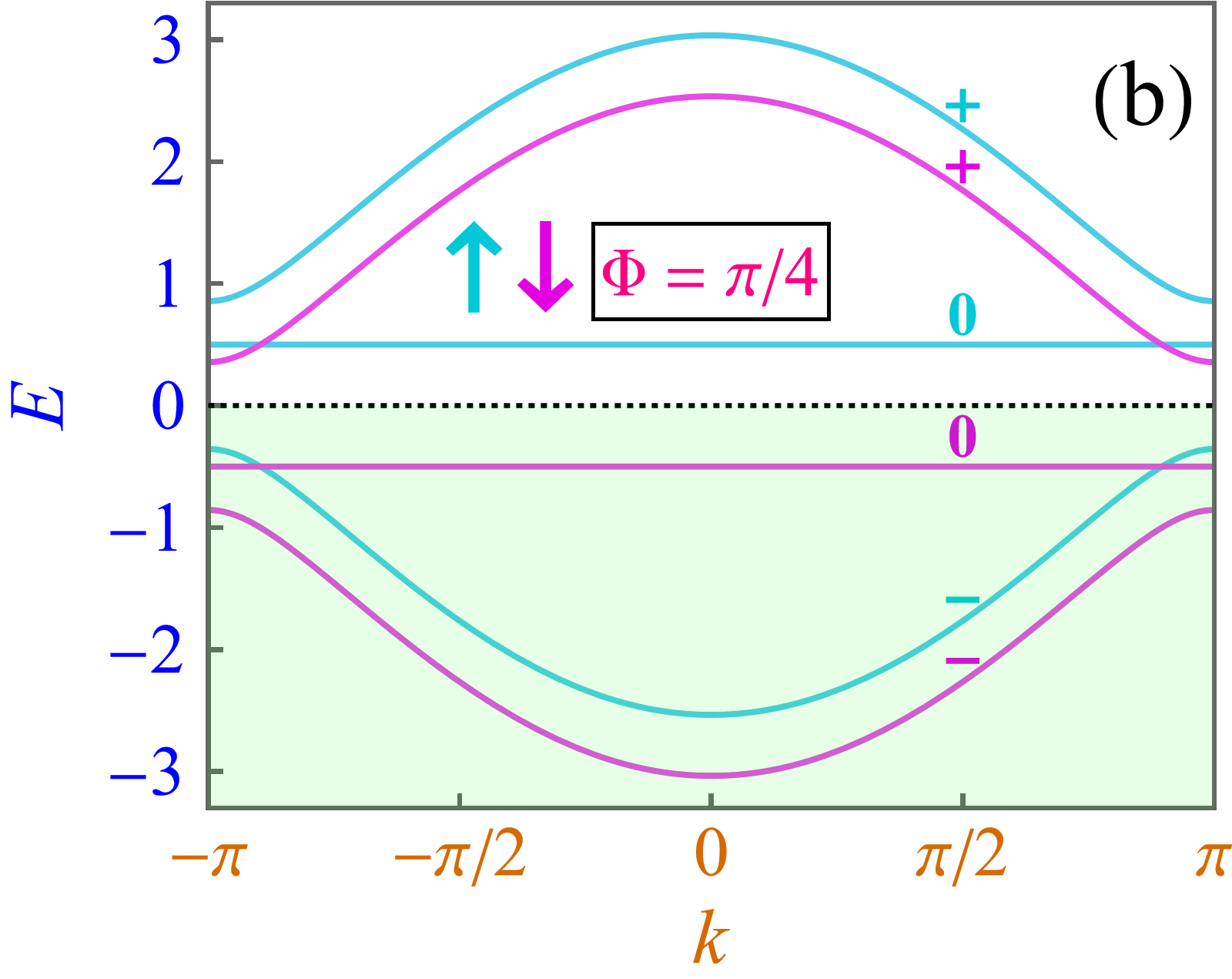}
\vskip 0.1cm
\includegraphics[clip, width=0.49\columnwidth]{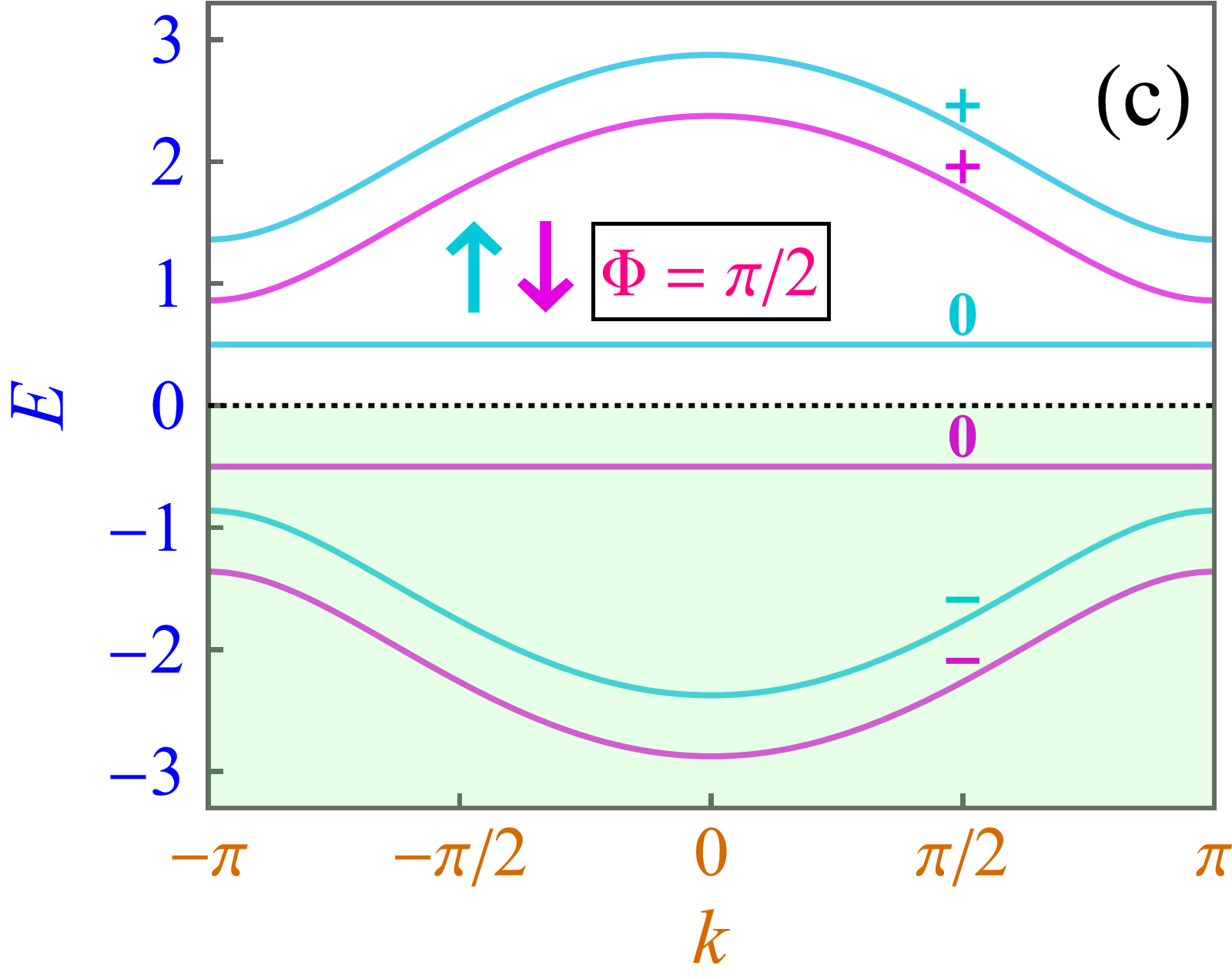}
\includegraphics[clip, width=0.49\columnwidth]{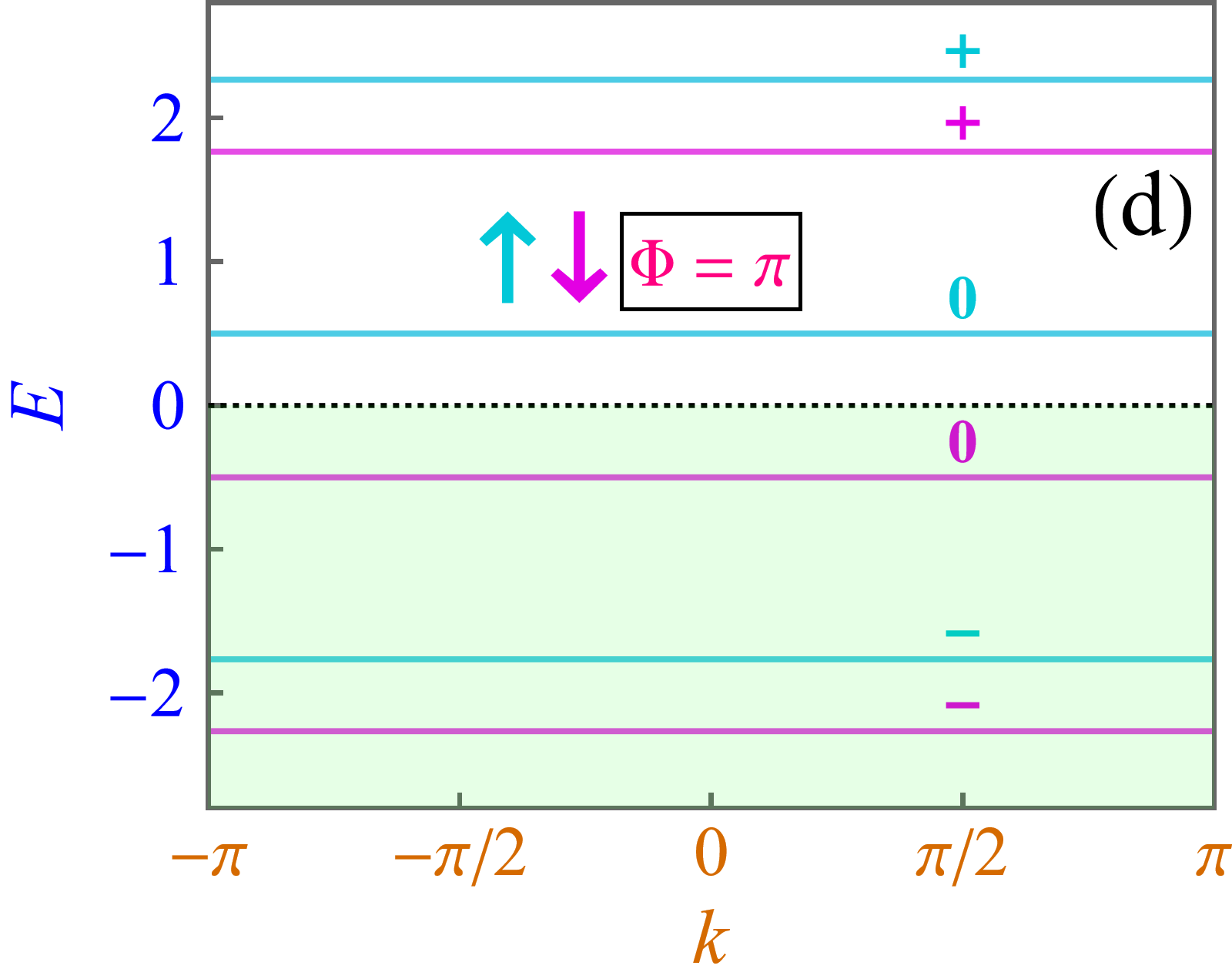}
\caption{Energy dispersions in both $\uparrow$ and $\downarrow$ sectors 
as a function of the momentum ($k$), for different values of the flux $\Phi$. 
The labels $+$, $-$, and $0$ correspond respectively to the two dispersive 
bands and to the  flat band of each spin sector (see main text). 
The green area indicates the occupied states. Here, we have chosen $JS=1$.}
\label{dispersion}
\end{figure}

In Fig.~\ref{dispersion}, we have depicted the energy dispersions for different values of the flux $\Phi$ and a fixed value of the local 
coupling $JS=1$. In what follows, we label the bands $(\sigma,l)$ where $l = +,-,\ \text{and}\ 0$. In the absence of flux, the bands 
$(\uparrow,0)$ and $(\uparrow,+)$ [resp. $(\downarrow,0)$ and $(\downarrow,-)$] exhibit a quadratic band touching at $|k|=\pi$. 
In addition, the bands $(\downarrow,+)$ and $(\uparrow,-)$ touch as well quadratically at the zone boundary of the Brillouin zone, 
which will have an effect on the nature of the couplings as discussed below. When the flux is switched on, in both spin sectors the 
flat band is now gapped, and a gap opens up between $(\downarrow,+)$ and $(\uparrow,-)$ at $|k|=\pi$. As $\Phi$ increases, the width 
of the dispersive bands reduces till all bands become completely flat when $\Phi=\pi$. As pointed out previously, this is the 
Aharonov-Bohm caging effect~\cite{Vidal-prl-2000,Vidal-prl-1998,Amrita-jpcm-2021}.

\section{Exchange couplings and magnetic force theorem}
\label{sec:Exchange-coupling}
To address the nature of the exchange couplings between the magnetic atoms in the flux-threaded magnetic diamond chain, we rely on the 
magnetic force theorem (MFT)~\cite{MFT1,MFT2,MFT3}. The MFT is a reliable and widely used tool in practical calculations of couplings 
in real compounds such as transition metals~\cite{pajda}, oxides~\cite{apl-zrO2}, Heusler compounds~\cite{heusler}, diluted magnetic 
semiconductors~\cite{RMP-DMS,GB-RB} and even in magnetic molecules~\cite{mazurenko} to name a few only. Within the MFT, the coupling 
between two spins located at $r_{i\lambda}$ and $r_{j\lambda'}$ is given by, 
\begin{equation}
J_{\lambda\lambda'}(R)=\frac{(JS)^{2}}{2}\int_{-\infty}^{+\infty} \chi_{ij}^{\lambda\lambda'}(\omega) f(\omega) d\omega, 
\label{eqcjij}
\end{equation}
where $R = r_{i\lambda} - r_{j\lambda'}$ and the generalized spin susceptibility $\chi_{ij}^{\lambda\lambda'}(\omega)=-\frac{1}{\pi}
\operatorname{\Im}\left[G_{ij\uparrow}^{\lambda\lambda'}(\omega) G_{ji\downarrow}^{\lambda'\lambda}(\omega) \right]$. The Green's function 
$\widehat{G}_{\sigma}(\omega)=(\omega+i\eta-\widehat{H}_{\sigma})^{-1}$, $\widehat{H}_{\sigma}$ being the Hamiltonian in the spin sector 
$\sigma$ ($\sigma =\: \uparrow,\downarrow$), $\eta$ mimics an infinitesimal inelastic scattering rate and 
$f(\omega)= \dfrac{1}{e^{(\omega-\mu)/k_{B}T}+1}$ is the Fermi-Dirac distribution. Here, we consider half-filled systems at $T=0\:K$ ($\mu=0$). 

Note that, the standard RKKY formula (second-order perturbation theory) is valid only in the perturbative regime, where 
$JS$ is smaller than the bandwidth. This breaks down when the Fermi energy lies in the FB. Indeed, applying the standard RKKY formula 
would lead to incorrect results, such as antiferromagnetic couplings and strong frustration effects as shown in 
Ref.~\cite{Georges-PRB-2021-RKKY}, while the non-perturbative calculation leads to strong ferromagnetic couplings and dominant contributions 
originating from the FB-FB contribution~\cite{mag-GB1}.

Eq.~\eqref{eqcjij} is derived for classical spins and implies that the corresponding effective Heisenberg Hamiltonian reads,
\begin{equation}
\mathcal{H}_{Heis}=\frac{1}{2}\sum_{i\lambda\ne j\lambda'} J_{\lambda\lambda'}(R)\,
{\bm e}_{i\lambda}\cdot{\bm e}_{j\lambda'}.
\label{Heisenberg}
\end{equation}
This implies that $J_{\lambda\lambda'}(R) \ge 0$ (resp. $J_{\lambda\lambda'}(R)\le 0$) means antiferromagnetic (resp. ferromagnetic) coupling.

It is very instructive to re-express the couplings in terms of the eigenstates of the Hamiltonian given in  Eq.\eqref{hamiltonian}, 
this will be useful for discussing the results of the following section. Eq.~\eqref{eqcjij} can be rewritten in terms of the 
different inter-band contributions as, 
\begin{equation}
J_{\lambda\lambda'}(R)=\sum_{pq} I_{\lambda\lambda'}^{pq}(R),
\label{eqjijb}
\end{equation}
where $p$ (resp. $q$) is the band index in the $\uparrow$ (resp. $\downarrow$) spin sector. The inter-band contribution reads, 
\begin{eqnarray}
I_{\lambda\lambda'}^{pq}(R)=\frac{(JS)^2}{2N^2} \sum_{k,k'}e^{i(k-k')R} A_{k\lambda,k'\lambda'}^{pq} \times  \nonumber \\ 
\frac{f(E^{\uparrow}_{p}(k))-f(E^{\downarrow}_{q}(k'))}{E^{\uparrow}_{p}(k)-E^{\downarrow}_{q}(k')},
\label{contrib}
\end{eqnarray}
$N$ is the number of cells, and the inter-band matrix element 
\begin{eqnarray}
 A_{k\lambda,k'\lambda'}^{pq} = 
 \langle k \lambda \hspace{-0.1cm} \uparrow \vert \psi^{\uparrow}_{p}(k) \rangle
\langle \psi^{\uparrow}_p(k) \vert k\lambda'\hspace{-0.1cm}\uparrow \rangle \cdot  \nonumber \\
\langle k'\lambda'\hspace{-0.1cm}\downarrow \vert \psi^{\downarrow}_{q}(k')\rangle
\langle \psi^{\downarrow}_{q}(k') \vert k'\lambda\hspace{-0.1cm}\downarrow \rangle,
\label{matrix-element}
\end{eqnarray}
where $\vert k\lambda\sigma \rangle =  \hat{c}_{k\lambda\sigma}^{\dagger} \vert 0\rangle$, $\hat{c}_{k\lambda\sigma}^{\dagger}$ 
being the Fourier transform of $\hat{c}_{i\lambda\sigma}^{\dagger}$. 
In what follows, we will focus on the AB flux dependence of certain specific contributions, such as the FB-FB contribution 
$I_{\lambda\lambda'}^{00}(R)$.

\section{Impact of the magnetic flux on the $(B,B)$ and $(B,C)$ couplings}
\label{sec:JBB-coupling}
\begin{figure}[ht]
\centering
\includegraphics[clip, width=0.95\columnwidth]{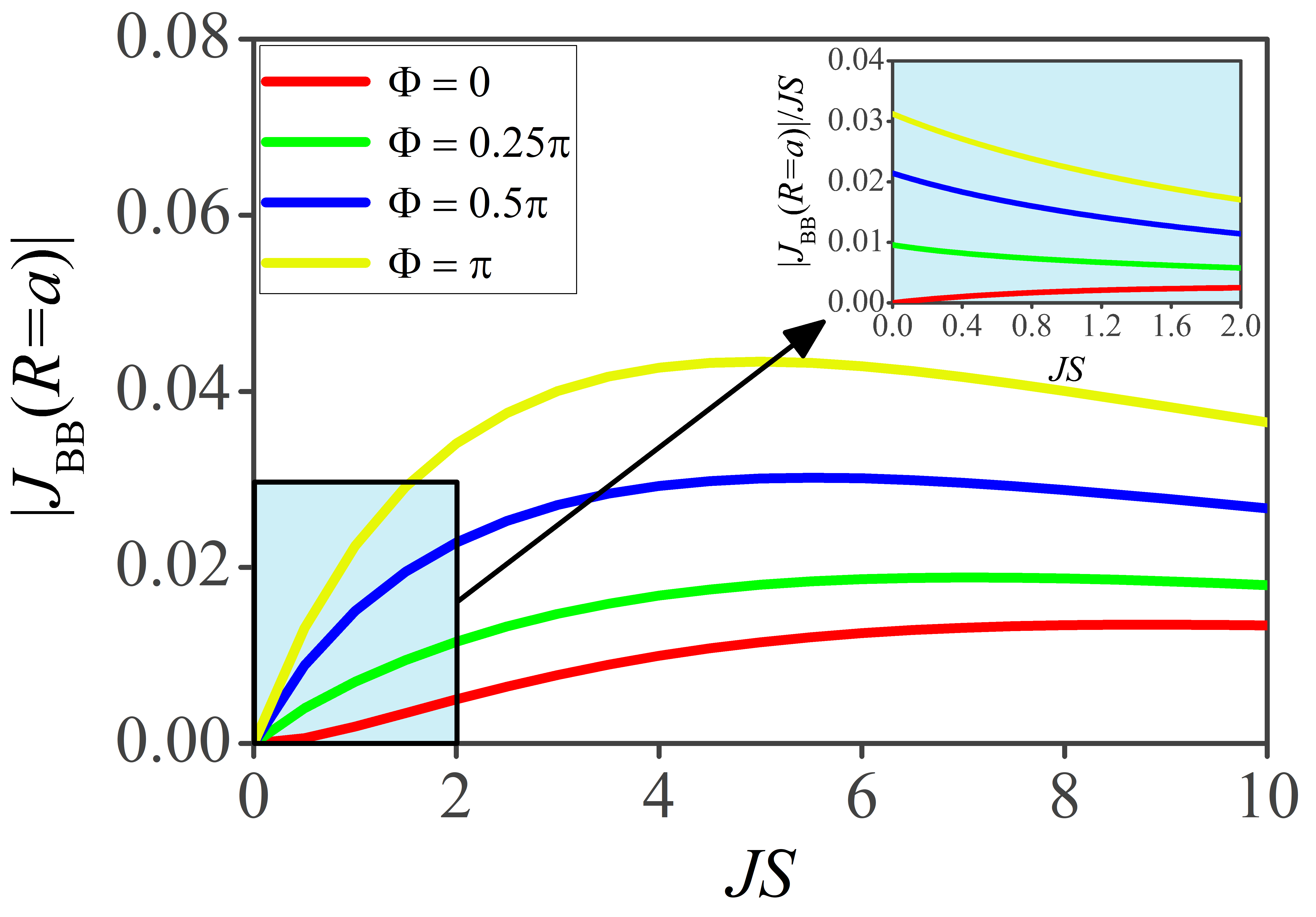}
\caption{Amplitude of the nearest-neighbor coupling $J_{BB}(R=a)$ as a 
function of $JS$ for different values of the magnetic flux $\Phi$, as 
depicted in the figure. The inset shows $J_{BB}(R=a)/JS$ as a function 
of $JS$ in the weak coupling regime.}
\label{JBB_vs_JS}
\end{figure}
In this section, we focus primarily on $J_{BB}(R)$, since by symmetry, $J_{CC}(R)$ is identical. Furthermore, we expect as well that 
$J_{BC}(R)=J_{BB}(R)$ for $R\ge a$. The only difference is the intra-cell coupling $J_{BC}(0)$, which is calculated separately.

In Fig.~\ref{JBB_vs_JS}, we have depicted the nearest-neighbor $(B,B)$ coupling as a function of $JS$ for the flux $\Phi$ ranging from 
$0$ to $\pi$. For small values of $JS$, we observe a rapid monotonic increase of the coupling as the flux increases. 
For instance, for $JS = 1$, the coupling jumps by an order of magnitude, from $J_{BB}(R=a) \sim 0.0025$ for $\Phi = 0$ to $J_{BB}(R=a) 
\sim 0.02$ for $\Phi = \pi$. In addition, as shown in the inset, $J_{BB}(R=a)$ scales quadratically with $JS$ when $\Phi=0$ and 
linearly (with an increasing slope) when $\Phi \ne 0$. As $JS$ increases further, $J_{BB}(R=a)$ continues to increase until it reaches 
a maximum whose position depends on the flux; for $\Phi=0$, $(JS)_{\textrm{max}} \approx 8t$, while for $\Phi=\pi$, $(JS)_{\textrm{max}} 
\approx 4t$. 

We now propose to shed some light on the change of behavior found in the regime of small $JS$ when a finite flux is introduced. First, 
we recall that in magnetic systems with FBs, the origin of couplings that vary linearly with $|JS|$ is the FB-FB contribution 
$I_{\lambda\lambda'}^{00}(R)$ which dominates in the weak coupling regime \cite{mag-GB1,mag-GB2}. We find, 
\begin{eqnarray}
I_{BB}^{00}(R) = -\frac{|JS|}{8N^2}\Big\vert
\sum_{k}\big[1-\frac{\sin(\frac{\Phi}{2})\sin(ka)}{1+\cos(\frac{\Phi}{2})\cos(ka)}\big]e^{ikR}
\Big\vert^2.
\label{I00}
\end{eqnarray}
In Appendix~\ref{AppendixA}, we provide the analytical expression of $I_{BB}^{00}(R)$. This equation shows that, when $\Phi=0$ this 
contribution vanishes and explains the origin of the quadratic scaling found in Fig.~\ref{JBB_vs_JS}. 
We return to this particular case in what follows. We remark that, Eq.~\eqref{I00} is valid for any value of $JS$, but 
this does not mean that coupling will diverge as $JS \rightarrow \infty$, this is discussed in what follows. 
As shown in Appendix~\ref{AppendixA}, for $R=a$, one finds $I_{BB}^{00}(R=a)=-\frac{|JS|}{8} {\big[}\frac{\sin(\Phi/2)}{1+\sin(\Phi/2)} {\big]}^2$.
\begin{figure}[ht]
\centering
\includegraphics[clip, width=0.95\columnwidth]{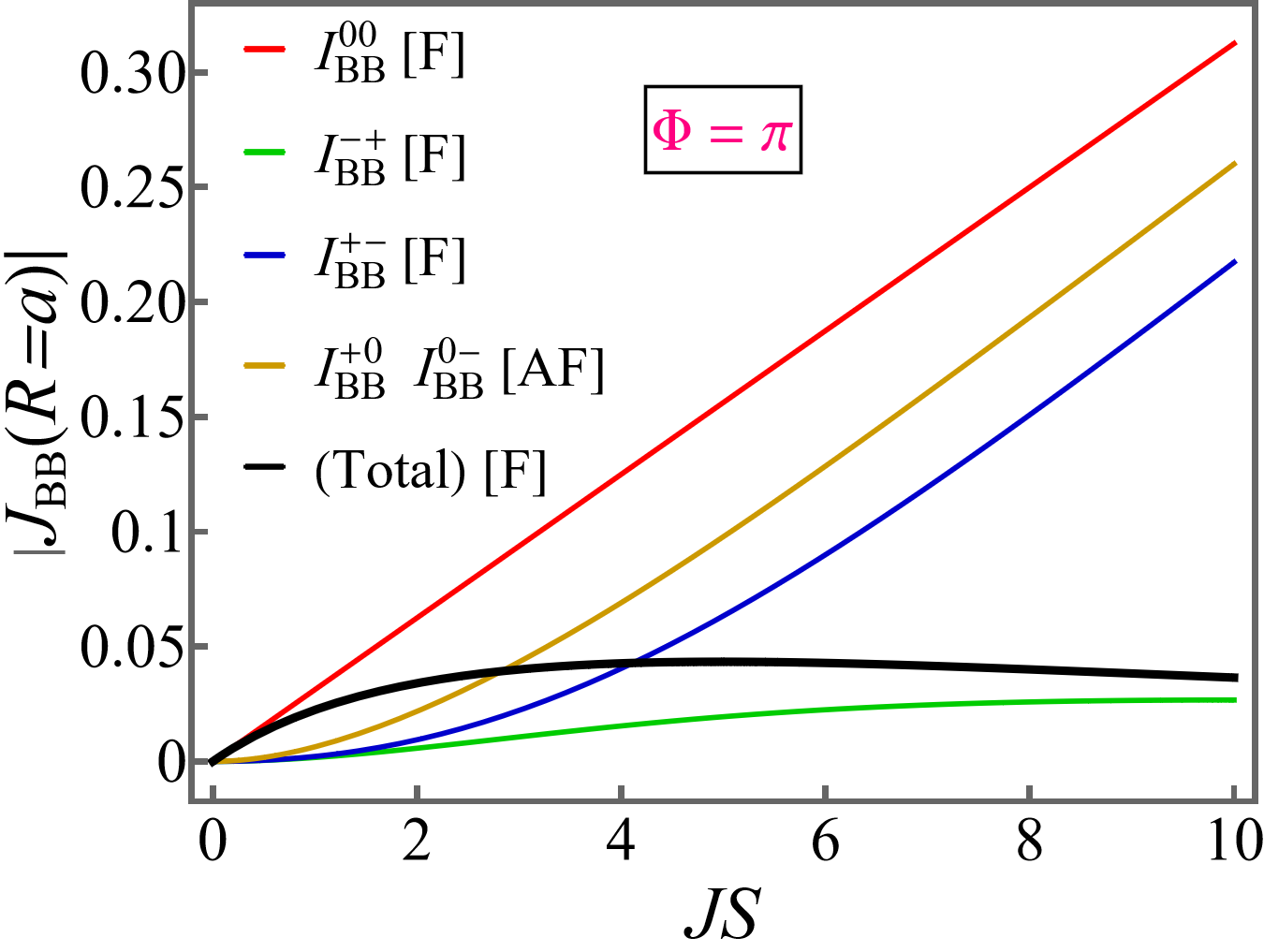}
\caption{Amplitude of the nearest-neighbor coupling between $B$ sites 
$J_{BB}(R=a)$ as a function of $JS$ (black line). The other lines 
correspond to the different contributions as discussed in the main text. 
`[F]' or `[AF]' means that the contribution is either ferromagnetic or 
antiferromagnetic. Here we have chosen $\Phi=\pi$.} 
\label{JBB_vs_JS_Pi-flux_Analytical}
\end{figure}

Let us now discuss the special case of $\pi$-flux threaded diamond plaquettes. All the bands are flat and the couplings 
can be straightforwardly calculated analytically (see Appendix~\ref{AppendixB}). 
For this value of $\Phi$, the couplings are restricted to the nearest-neighbor only. 
The individual contributions to the couplings are depicted in Fig.~\ref{JBB_vs_JS_Pi-flux_Analytical}. All terms are 
ferromagnetic (F), except $I_{BB}^{+0}$ and $I_{BB}^{0-}$ which are antiferromagnetic (AF). In the limit of large $JS$, 
the AF contributions exactly compensate the $I_{BB}^{00}$ and $I_{BB}^{+-}$ leading to a $1/JS$ decay of the full coupling 
and $J_{BB} = I_{BB}^{-+}$ (see Fig.~\ref{JBB_vs_JS_scaling} in Appendix ~\ref{AppendixB}). We remark that, the strong 
increase of the nearest-neighbor coupling when $\Phi$ varies from $0$ to $\pi$ (in the weak coupling regime $JS<t$) can be 
understood as the result of the combined effects of (\textit{i}) modification of the wavefunction overlap, (\textit{ii}) 
increase of the localization effect (AB caging), and (\textit{iii}) redistribution of the spectral weight between the flat 
and the dispersive bands.
\begin{figure}[ht]
\centering
\includegraphics[clip, width=0.95\columnwidth]{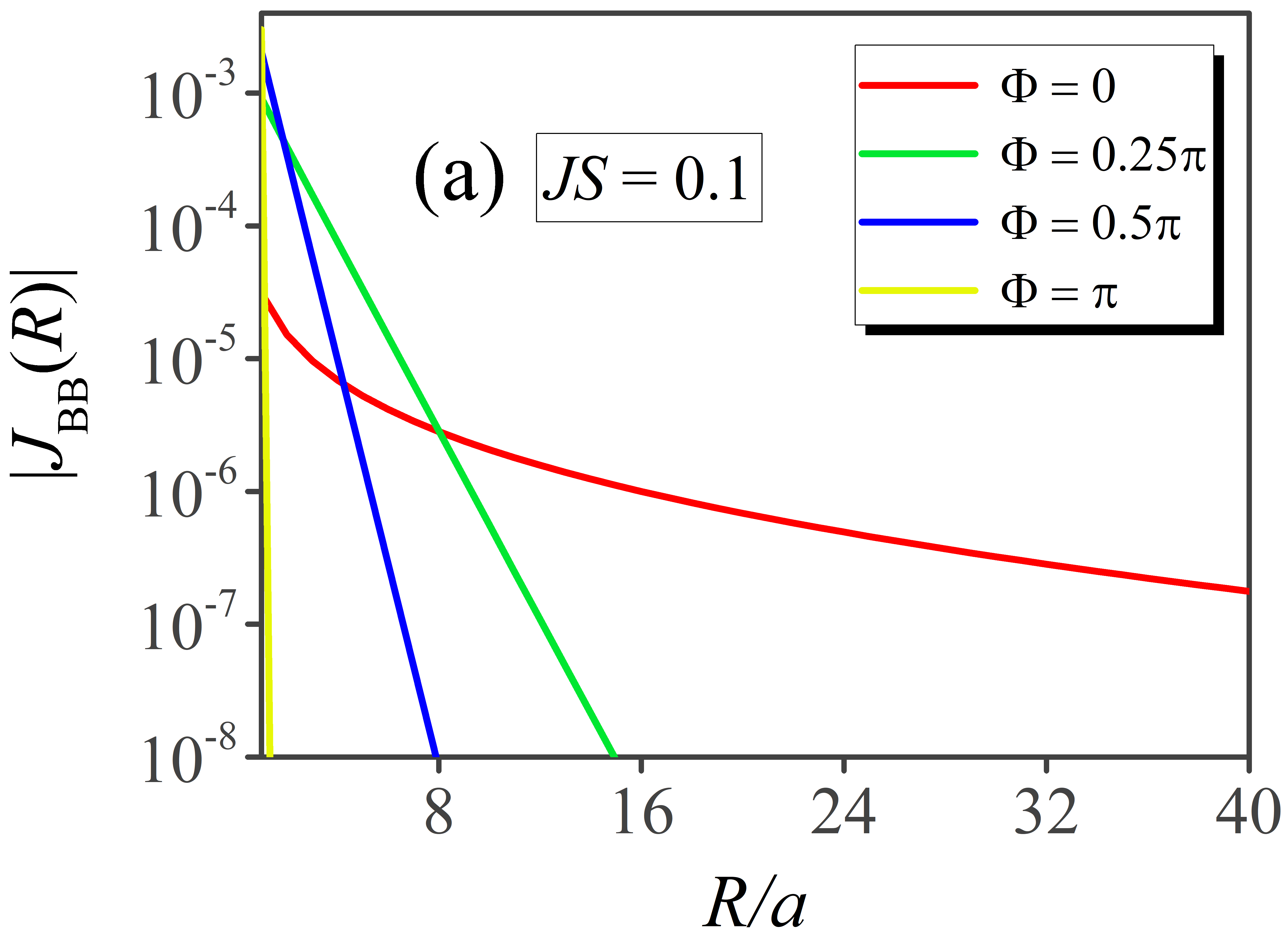}
\vskip 0.1cm
\includegraphics[clip, width=0.95\columnwidth]{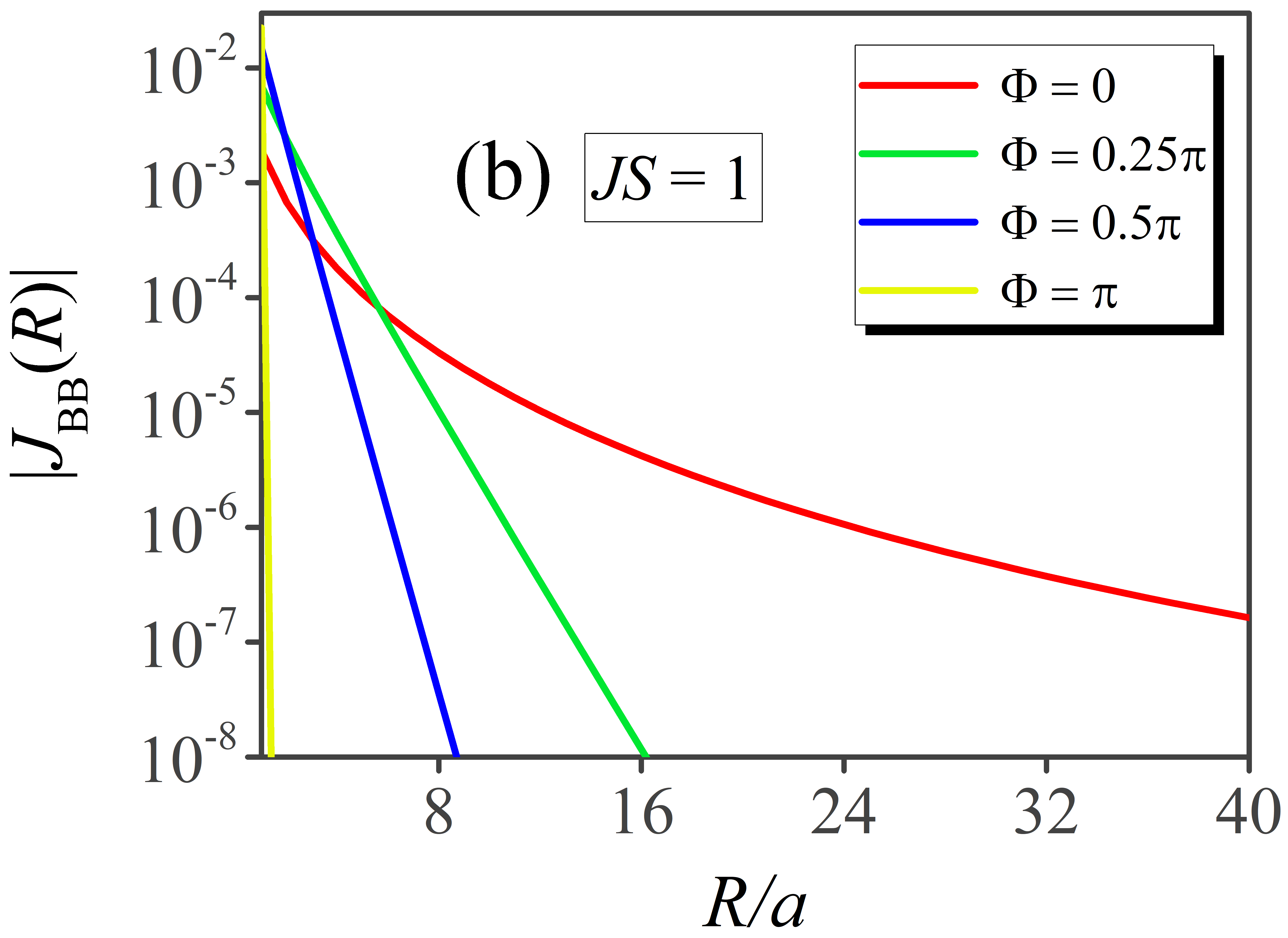}
\caption{Coupling $J_{BB}(R)$ as a function of $R$ for (a) $JS=0.1$ and 
(b) $JS=1$. The values of the flux $\Phi$ associated to each color are 
depicted in both the panels.}
\label{JBB_vs_R}
\end{figure}

We now propose to discuss the impact of the flux on the decaying length of the magnetic coupling. In Fig.~\ref{JBB_vs_R}(a) 
and (b), we have plotted $\vert J_{BB}(R)\vert$ as a function of the distance $R$ for $\Phi$ ranging from $0$ to $\pi$. 
We have considered two different values of $JS$; $JS =0.1$ in panel (a) and $JS = 1$ in panel (b). First, qualitatively, 
we observe a similar behavior; the coupling scales as $e^{-R/\xi(\Phi)}$, where $\xi$ reduces as the flux increases. 
As pointed out previously, when $\Phi=\pi$, the couplings are restricted to the nearest-neighbor only, which means that 
a fit gives $\xi=0$. For these values of $JS$, when $\Phi \ne 0$ the coupling is dominated by the FB-FB term; thus 
$J_{BB}(R) \approx I_{BB}^{00}(R)$ whose analytical expression is given in Appendix~\ref{AppendixA}.

The special case of $\Phi=0$ was addressed in Ref.~\cite{mag-GB2}. Among the contributions $I_{BB}^{pq}(R)$, only one involves 
matrix elements between two touching bands: $p = -$ and $q = +$ (see Fig.~\ref{dispersion}). This is the dominant contribution 
to $J_{BB}(R)$, which is expected to decrease according to a power law $1/R^{\alpha}$. In  Ref.~\cite{mag-GB2}, it was 
analytically demonstrated that $J_{BB}(R) = -\frac{3t^2}{2\pi|JS|} \frac{1}{R^4}$, which is valid for any value of $JS$, under 
the condition that $R \gg \sqrt{32}\pi at/|JS|$.
\begin{figure}[ht]
\centering
\includegraphics[clip, width=0.95\columnwidth]{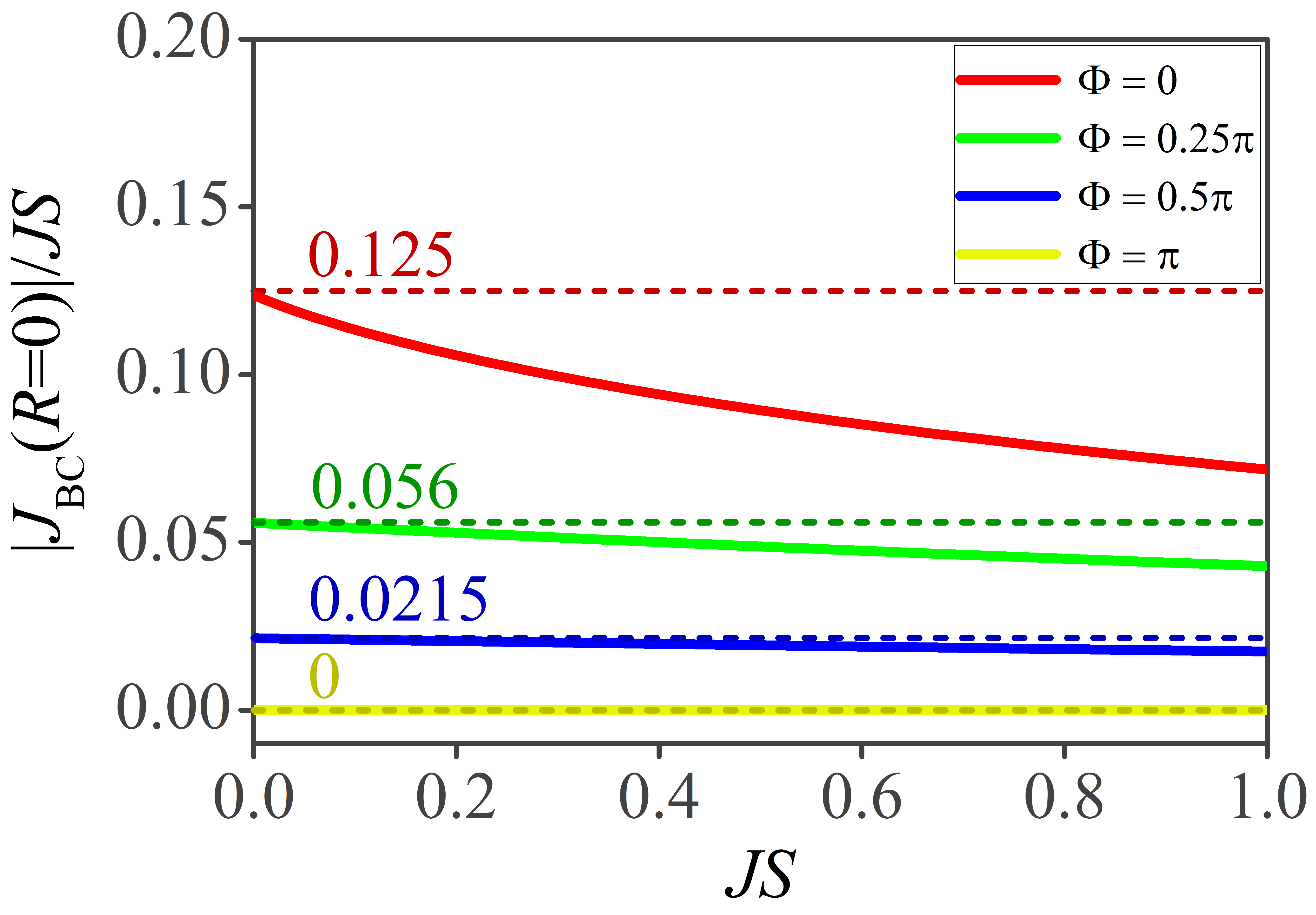}
\caption{Rescaled intra-cell $BC$-coupling  $J_{BC}(R=0)$ as a function 
of $JS$ for different values of the flux $\Phi$. The horizontal dashed 
lines correspond to the analytical values for $JS \rightarrow 0$, which 
correspond to the FB-FB contribution $\vert I_{BC}^{00}(R=0)/JS\vert$ 
(see main text).}
\label{JBC_vs_JS}
\end{figure}

We now discuss the effect of $\Phi$ on the $(B,C)$ coupling and focus on the weak coupling regime. As pointed out 
above, $J_{BC}(R)=J_{BB}(R)$ for $R\ge a$. However, the intra-cell $(B,C)$ coupling ($R=0$) requires a separate 
calculation. In Fig.~\ref{JBC_vs_JS}, we have depicted $J_{BC}(0)$ as a function of $JS$ for $\Phi/\pi=0$, $1/4$, 
$1/2$ and $1$. First, for a given value of $JS$, we observe a decrease in $J_{BC}(0)$ as $\Phi$ increases, which 
contrasts with what was found for $J_{BB}(a)$. We remark that, for the specific case of $\pi$ flux in each plaquette, 
we find $J_{BC}(0)=0$ for any value of $JS$. Secondly, we find that for any value of $\Phi$, the coupling scales 
linearly with $JS$ even for $\Phi=0$, which indicates that in the limit of small $JS$, $J_{BC}(0)$ is dominated 
by the FB-FB contribution. However, the deviation from the linear behavior increases strongly as the flux approaches 
to $0$, which shows that the other contributions to the coupling rapidly contribute. The expression for the FB-FB 
contribution can be obtained analytically. One finds,
\begin{equation}
I_{BC}^{00}(0)=-\frac{|JS|}{8} \left[\frac{1-\sin(\Phi/2)}{1+\sin(\Phi/2)}\right]. 
\label{I_BC_00_analytical}
\end{equation}
The intra-cell FB-FB contribution decays as the flux increases and vanishes when $\Phi=\pi$.  

\section{Magnon spectrum and magnon thermal conductivity}
\label{sec:magnon-spectrum}
In this section, we first discuss how the magnon spectrum is affected by the AB flux. The magnetic system, consisting 
of two localized spins per unit cell, has two distinct magnon branches: one is optical (gapped mode) and the other is 
acoustic (gapless mode). Starting with the Heisenberg Hamiltonian $\mathcal{H}_{Heis}$ as defined in Eq.~\eqref{Heisenberg} 
and following the equation of motion method (RPA decoupling)~\cite{tyablicov,gb-pb}, one finds that the eigenmodes in 
the magnetic diamond chain are the solution of $det[\mathcal{D}(\omega)]=0$, where the $2\times2$ matrix,
\begin{align}
\mathcal{D}(\omega)= 
\left[\def\arraystretch{1.5} \begin{matrix}
\omega - Sf_{BB}(k) & Sf_{BC}(k) \\
Sf_{CB}(k) & \omega - Sf_{CC}(k) \\
\end{matrix}\right],
\label{magnon-modes}
\end{align}
%
with $f_{BB} (k) = f_{CC}(k)= \bar{J}_{BB}(k)-2\bar{J}_{BB}(k=0)-J_{BC}(R=0)$ and $f_{BC}(k)=f_{CB}(k) = -\bar{J}_{BB}(k)+J_{BC}(R=0)$, 
with $\bar{J}_{BB}(k)=\sum_{R} e^{ikR} J_{BB}(R)$. Note that, we have used the fact that $J_{BB}(R)=J_{CC}(R)=J_{BC}(R)$ for $R \ne 0$.
We find two magnon branches, one being dispersive and the second flat: $\omega_{1,2}(k)=S E_{1,2}(k)$ where, 
\begin{subequations}
\begin{align}
E_1(k) & = -2\Big[
\bar{J}_{BB}(k=0)-\bar{J}_{BB}(k)\Big], \\
E_2(k)& = -2\Big[\sum_{R}J_{BB}(R)+J_{BC}(R=0)\Big].    
\end{align} 
\end{subequations}

\begin{figure}[ht]
\centering
\includegraphics[clip, width=0.95\columnwidth]{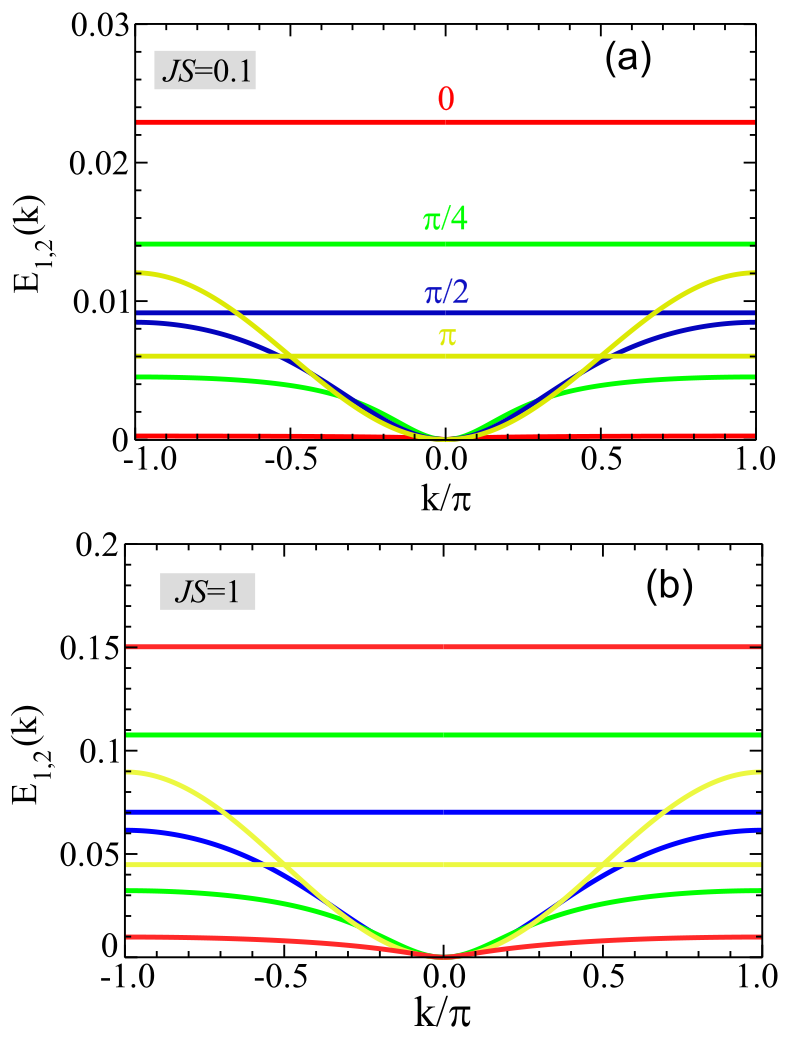}
\caption{Magnon modes $E_1$ (dispersive mode) and $E_2$ (flat mode) 
as a function of $k$ for both (a) $JS=0.1$ and (b) $JS=1$. The 
values of the flux $\Phi$ which correspond to different colors are 
depicted in the panel (a).}
\label{magnon-modes}
\end{figure}

As expected one can check that $E_{1}(k)= D_{S} k^2$ in the limit of vanishing momentum (Goldstone mode). The impact of the flux 
on the magnon modes is depicted in Fig.~\ref{magnon-modes}(a) and (b) for two different values of $JS$. In both cases, we 
observe that as $\Phi$ increases, the energy of the flat mode reduces. On the other hand, the dispersive mode behaves 
differently. As $\Phi$ is switched on, we find a rapid and strong increase of the magnon bandwidth. The ratio between 
the bandwidth at $\Phi=\pi$ divided by that at $\Phi=0$ is approximately $9$ for $JS = 1$. It is more impressive when 
$JS = 0.1$, since the ratio reaches $50$. This huge value reflects the fact that in the weak coupling regime, the couplings 
are largely dominated by the FB-FB contribution, as discussed in the previous section. 

We now propose to estimate the impact of an AB flux on the thermal transport by magnons. With the spirit of keeping 
things simple, we consider the constant relaxation time approximation, where the energy and temperature dependence of 
the magnons is ignored. Similar to the phonon thermal conductivity, that of magnons is given by, 
\begin{equation}
\kappa (T) =\frac{1}{2\pi} \int_{-\pi}^{+\pi} dk\: v^2(k)\omega(k) \tau \frac{\partial f_{BE}}{\partial T},
\label{kappa}
\end{equation}
where $v(k)=\frac{1}{\hbar}\frac{\partial\omega(k)}{\partial k}$ is the magnon velocity, $\tau$ is the relaxation time 
of the magnon mode and $f_{BE}=\left[{e^{\omega(k)/k_{B}T}-1}\right]^{-1}$ is the Bose-Einstein distribution. 
Eq.~\eqref{kappa} can be derived following the well-known standard Boltzmann transport equation method.
\begin{figure}[ht]
\centering
\includegraphics[clip, width=0.95\columnwidth]{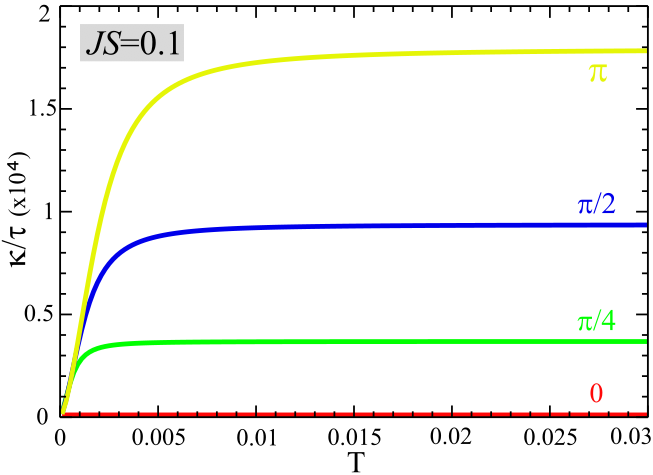}
\caption{Magnon thermal conductivity (rescaled) in the constant relaxation 
time approximation as a function of temperature $T$ (in unit of $t$). The 
colors correspond to different values of $\Phi$ as in Fig.~\ref{magnon-modes}(a). 
In the case $\Phi=0$, $\kappa$ is quasi negligible. Here, we have chosen $JS=0.1$.}
\label{thermal-conductivity}
\end{figure}

In Fig.~\ref{thermal-conductivity}, $\kappa$ is plotted as a function of the temperature $T$, for $JS=0.1$ and 
different values of $\Phi$. First, we immediately see that  $\kappa$ at $\Phi=0$ is negligible. More precisely, it is 
about 3 orders of magnitude smaller than that obtained when $\Phi=\pi$. In addition, as the flux varies from 
$\Phi = \pi/4$ to $\Phi=\pi$, one finds a boost of $\kappa$, which increases by about $500\%$ (for $T \ge 0.005\,t$). 
We believe that, these findings could be of interest for energy conversion technologies and nanoscale thermal transport, 
which is relevant for quantum information processing and development of magnonic devices~\cite{Uchida2021,Pirro2021,Baumgaertl2023}. 
Note that, the relaxation time approximation has the great advantage of leading to a simple estimate of the 
thermal conductivity, but one should keep in mind that it provides more of a qualitative than a quantitative estimate. 
In-depth calculations go beyond the scope of the present study.

\section{Connection between the decaying length and the quantum metric}
\label{sec:quantum-metric}
Previously, we have found that for small values of $JS$ and $\Phi \ne 0$, $J_{BB}(R) \approx I_{BB}^{00}(R)$, whose analytical 
expression is given in Appendix~\ref{AppendixA}. In this regime of weak coupling, we can extract the decaying length and find, 
\begin{equation}
\xi(\Phi)=\frac{a}{2\big|\vert\ln{z_+|}\big|}
\label{xi}
\end{equation} 
with $z_{+}=\frac{1}{\cos(\Phi/2)}\left[-1 + \sin(\Phi/2)\right]$ for $0 < \Phi < \pi$. 
Thus, when $\Phi \rightarrow 0$ (resp. $\Phi \rightarrow \pi$), the decaying length diverges (resp. vanishes) and 
$\xi \simeq a/\Phi$ (resp. $\xi \simeq a/4\vert \ln{(\pi- \Phi)}\vert$. 
\begin{figure}[ht]
\centering
\includegraphics[clip, width=0.95\columnwidth]{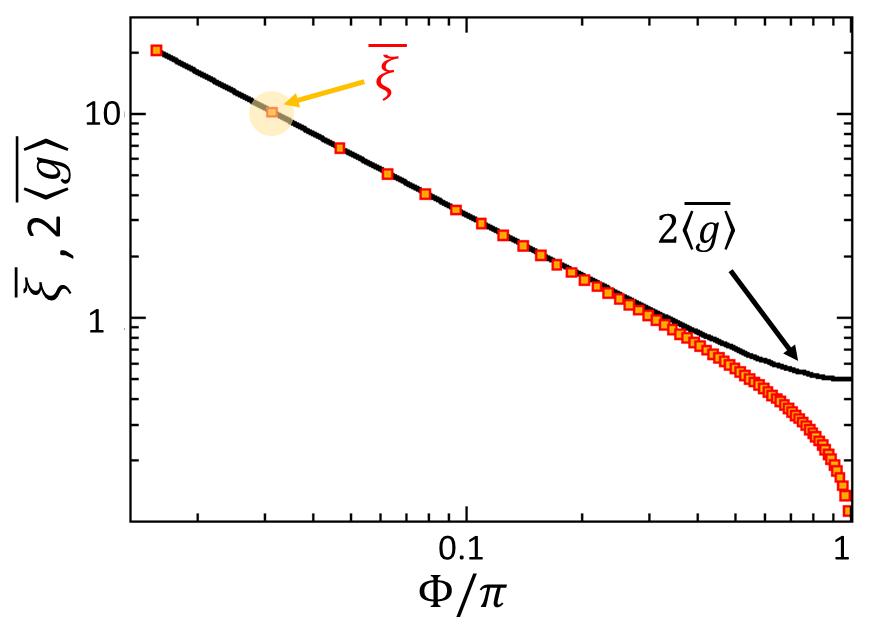}
\caption{$\overline{\xi}=\xi (\Phi)/a$ (continuous black line) and 
$2\overline{\langle g \rangle}=2\langle g \rangle/a^2$ (square red symbols) 
as a function of the flux $\Phi$.}
\label{decaying-length}
\end{figure}
In a recent study~\cite{mag-GB2} on dilute magnetic stub chains, it was shown that the characteristic distance over which 
couplings decay can depend on the quantum metric of the flat band eigenstates in a complex way. We briefly recall that, 
the quantum metric (QM) corresponds to the real part of the quantum geometric tensor~\cite{qm1,qm2,qm3}, 
while its imaginary part is associated with the Berry curvature. The QM is defined as,
\begin{equation}
g(k) = \langle \partial_k \Psi^{0}_k \vert \partial_k \Psi^{0}_k \rangle - 
\vert \langle \partial_k \Psi^{0}_k \vert \Psi^{0}_k \rangle \vert^2, 
\label{quantum-metric}
\end{equation} 
where $\vert \Psi^{0}_k \rangle$ is the FB eigenstate at $JS=0$. It should be noted that the average of the QM over the 
Brillouin zone can be interpreted as a measure of the typical spreading of the FB eigenstates.

It is interesting to compare $\xi(\Phi)$ to the average value of the QM associated to the FB eigenstates. Let us define
$\langle g \rangle = \frac{1}{2\pi}\int_{-\pi}^{+\pi} g(k)dk$, the average value of the QM. Using the expression of the 
FB eigenstate one finds, 
\begin{equation}
g(k)=\frac{a^2\sin^2(\frac{\Phi}{2})}{4\left[1+\cos(\frac{\Phi}{2}) \cos(ka)\right]^2}. 
\end{equation}
After performing the integration over the momentum $k$ one obtains,
\begin{equation}
\langle g \rangle = \frac{a^2}{4|\sin(\frac{\Phi}{2})|}.
\label{QM}
\end{equation}

Thus, in the limit of vanishing flux, the QM diverges. This point deserves somehow a small discussion. Indeed, the FB eigenstate 
at $\Phi=0$ is momentum independent since the support of its CLS is contained in a single unit cell. This immediately implies that 
$ g(k) = 0$ and hence $\langle g \rangle = 0$, which contradicts Eq.~\eqref{QM}. The origin of the conflict is the fact that, 
at $\Phi=0$, the QM is not well-defined because of the band touching at $|k| = \pi$ as illustrated in Fig.~\ref{dispersion}(a). 

In Fig.~\ref{decaying-length}, we have depicted the quantum metric and $\xi$ as a function of the flux $\Phi$. An interesting 
result is revealed -- the decaying length coincides with the quantum metric when the flux is small enough. More precisely, when 
$\Phi \le 0.3 \pi $, we find with a good numerical accuracy that $\xi=2\langle g\rangle/a$. On the other hand, as the 
flux $\Phi$ approaches the regime of AB caging (\ie, $\Phi = \pi)$, we observe that $\xi$ and $\langle g \rangle$ behave differently; 
while $\xi \rightarrow 0$, the QM saturates, \eg, $\langle g \rangle \rightarrow 1/4$.

\section{Robustness against perturbations}
\label{sec:perturbation-in-J}
\begin{figure}[ht]
\centering
\includegraphics[clip, width=0.95\columnwidth]{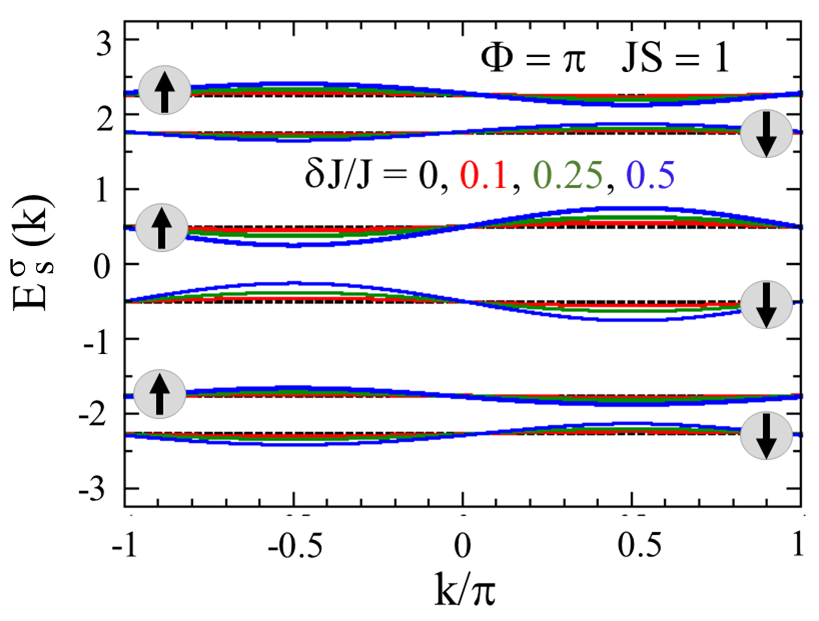}
\caption{Energy dispersions $E^{\sigma}_{s}$ as a function of momentum, where 
$\sigma =\uparrow,\:\downarrow$ (depicted in the figure) and $s = 0,\pm$ and 
several values of the ratio $\delta J/J$ corresponding to different colors. 
Here, we have taken $\Phi=\pi$ and $JS = 1$.}
\label{dispersion-with-perturbation}
\end{figure}

So far we have essentially considered the case where the FBs are rigorously flat. However, in real compounds, the hoppings 
or/and the local couplings will very likely be subject to possible perturbations. Hence, it is interesting and relevant to 
consider in this section the situation, where the flat bands become weakly dispersive. To study how a finite bandwidth of 
the flat bands affects our findings, we propose to break the equivalence between the way the $B$ and $C$ orbitals are coupled 
to the localized spins; for this, we replace the second term of the Hamiltonian~\eqref{hamiltonian} as follows,
\begin{eqnarray} 
J\sum_{i,\lambda \in \Lambda'}\hat{\bm s}_{i\lambda}\cdot {\bm S}_{i\lambda} \Rightarrow 
(J+\delta J)\:\hat{\bm s}_{i B}\cdot {\bm S}_{i B} + \nonumber \\ 
(J-\delta J)\:\hat{\bm s}_{i C}\cdot {\bm S}_{i C}.
\label{hamiltonian-modified}
\end{eqnarray}

For the sake of clarity, in this section, we consider the case where local coupling $JS = 1$, which will not have an impact on 
our conclusion. First, as an illustration, in Fig.~\ref{dispersion-with-perturbation}, we have plotted the dispersions for several 
values of $\delta J/J$ and $\Phi = \pi$. As can be seen, in this case, all the flat bands become dispersive, with a weaker impact 
in the vicinity of $|k| = 0,\pi$. As expected, the width of the quasi FBs increases almost linearly with $\delta J$.
\begin{figure}[ht]
\centering
\includegraphics[clip, width=1.05\columnwidth]{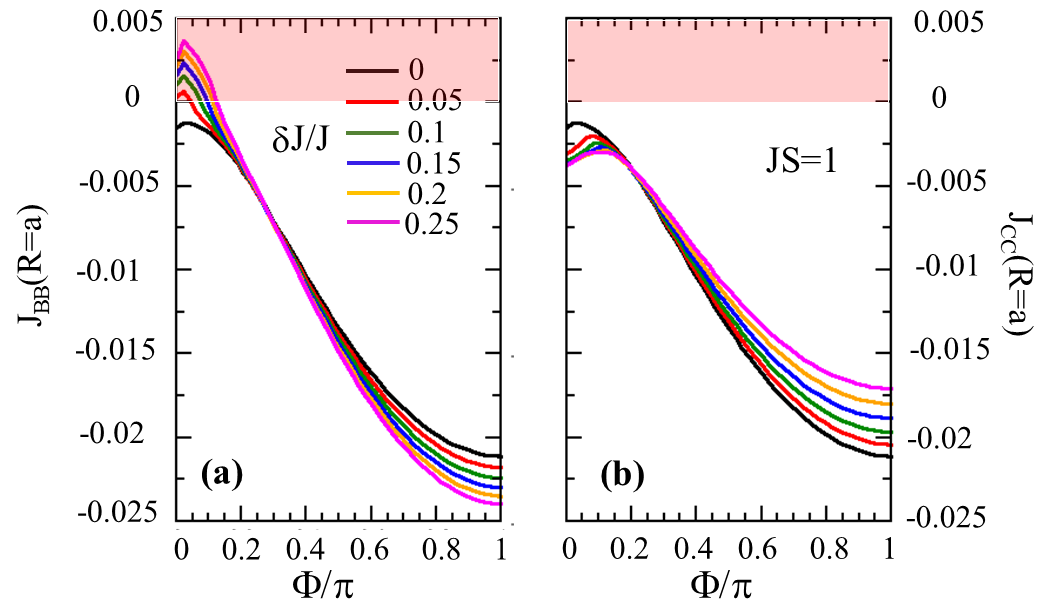}
\caption{(a) $J_{BB}(R=a)$ and (b) $J_{CC}(R=a)$ as a function of 
$\Phi$ for different values of $\delta J$ and $JS=1$ (see main text). 
The red area corresponds to antiferromagnetic couplings.}
\label{JBB-with-perturbation}
\end{figure}

Fig.~\ref{JBB-with-perturbation}(a) and (b) represents  respectively the $(B,B)$ and $(C,C)$ couplings between nearest-neighbor 
pairs as a function of the flux for several values of $\delta J$. We consider first $J_{BB}(R=a)$. For small values of $\Phi$, 
we find a rapid suppression of the ferromagnetic character of the coupling. Indeed, for $\delta J \ge 0.05$, $J_{BB}(R=a)$
becomes antiferromagnetic. In contrast, in the vicinity of $\Phi=\pi$, $\delta J$ has the opposite effect since the ferromagnetic 
coupling is enhanced. On the other hand, $J_{CC}(R=a)$ remains ferromagnetic at small flux and its amplitude increases as 
$\delta J$ is switched on. For a flux $\Phi \approx \pi$, we observe that $J_{CC}(R=a)$ decreases only slightly as $\delta J$ 
increases. More precisely, we find that $\delta J_{CC} = -0.017 \: \delta J$. Thus, Fig.~\ref{JBB-with-perturbation} confirms 
that the strong increase in the couplings induced by the AB flux seems to be robust against the perturbations which make the 
FBs weakly dispersive. We have found that our results are robust against strong perturbation of the local couplings. This can be 
explained by the fact that the electronic bands remain quasi-flat and their effective bandwidths are always much smaller than 
the inter-band gaps.

\section{Conclusions}
\label{sec:conclu}
In conclusion, this work has shown that exchange couplings at short distances in the magnetic diamond chain can be strongly 
amplified in the presence of an AB flux and also lead to a giant increase in the magnon thermal conductivity. The effect is 
particularly pronounced when the local coupling between the itinerant carrier and the localized spins is weak. It is also 
revealed that, this phenomenon is robust against perturbations that would make the flat bands weakly dispersive. We have 
established a connection between the flux-dependent decaying length of the couplings and the quantum metric of the flat bands. 
Although, the present study focused only on the case of the diamond chain, our results are general and can be applied to 
other geometries and lattices exhibiting flat bands in the vicinity of the half-filled system. We believe that our results
could be of interest for both spintronics and magnonics.
%
\begin{acknowledgments}
BP gratefully acknowledges the local hospitality and the funding provided by 
the CNRS during his research visit to the N\'{e}el Institute, where this work 
was initiated. BP would also like to thank the Nagaland University for providing 
a minor start-up research grant. 
\end{acknowledgments}
%
\appendix
\section{Flux dependence of the FB-FB contribution to the $(B,B)$ coupling} 
\label{AppendixA}

In this appendix, we propose to derive the analytical expression of $I^{00}_{BB}$, the FB-FB contribution to the $(B,B)$ coupling. 
Assuming a ferromagnetic spin texture for the localized spins, in the basis $(\hat{c}_{kA\sigma}^{\dagger},\: \hat{c}_{kB\sigma}^{\dagger},\: 
\hat{c}_{kC\sigma}^{\dagger})$ in $k$ space, the Hamiltonian given in equation Eq.~\eqref{hamiltonian} becomes, 
\begin{eqnarray}
\widehat{H} = 
\begin{bmatrix}
\widehat{h}_{\uparrow}  & 0_{3\times3}\\
0_{3\times3} & \widehat{h}_{\downarrow}\\
\end{bmatrix},
\end{eqnarray}
where,
\begin{eqnarray}
\widehat{h}_{\sigma} =
\begin{bmatrix}
0  & f_{+} & f_{-}\\
f_{+} & \epsilon_{\sigma}  & 0\\
f_{-} & 0 & \epsilon_{\sigma}
\end{bmatrix},
\end{eqnarray}
with $f_{\pm} = -2t\cos(k/2\pm\Phi/4)$,
$\epsilon_{\sigma}= z_{\sigma}\frac{JS}{2}$ and $z_{\sigma} = \pm 1$ for $\sigma = \uparrow$,\: $\downarrow$. 
In each spin sector, the energy spectrum reads,
$E^{\sigma}_{\pm}=\frac{1}{2}\big[z_{\sigma}\frac{JS}{2} \pm \sqrt{\Delta_\Phi(k)}\big]$ and 
$E^{\sigma}_{0}=z_{\sigma}\frac{JS}{2}$ where $\Delta_{\Phi}(k)=\frac{1}{4}J^2S^2 + 16t^2\big[1+\cos(\frac{\Phi}{2})\cos(k)\big]$. 
The corresponding eigenvectors are $\vert \Psi^{\sigma}_{\pm}\rangle=
\frac{1}{\sqrt{N^{\sigma}_{\pm}}}(E^{\sigma}_{\pm}-z_{\sigma}\frac{JS}{2},\: -f_{-},\: -f_{+})$, 
where $N^{\sigma}_{\pm}=4t^2\big[1+\cos(\frac{\Phi}{2})\cos(k)\big] + \big[E^{\sigma}_{\pm}-z_{\sigma}\frac{JS}{2}\big]^2$. 
The FB eigenstate has the same expression in both spin sectors and is given by 
$\vert \Psi^{\sigma}_{0}\rangle = \frac{1}{\sqrt{1+\cos(\frac{\Phi}{2})\cos(k)}}(0,f_{+}/2,-f_{-}/2)$.

The FB-FB contribution to the $(B,B)$ coupling is given by, 
\begin{eqnarray}
I_{BB}^{00}(R)=\frac{(JS)^2}{2N^2}\sum_{k,k'}e^{i(k-k')R} A_{kB,k'B}^{00} \times  \nonumber \\ 
\frac{f(E^{\uparrow}_{0}(k))-f(E^{\downarrow}_{0}(k'))}{E^{\uparrow}_{0}(k)-E^{\downarrow}_{0}(k')},
\end{eqnarray}
with $A_{kB,k'B}^{00} = |\langle kB\vert \Psi^{\uparrow}_{0} \rangle|^2  \times |\langle k'B\vert \Psi^{\downarrow}_{0} \rangle|^2$.
This becomes 
\begin{eqnarray}
I_{BB}^{00}(R)=-\frac{|JS|}{2} \vert F^{BB}_{\Phi}(R) \vert^2,
\end{eqnarray}
where for simplicity, we have introduced $F^{BB}_{\Phi}(R)=\frac{1}{N}\sum_{k}e^{ikR}\frac{\cos^2(k/2+\Phi/4)}{\left[1+\cos(\Phi/2)\cos(k)\right]}$. 
It can be re-written as, 
\begin{eqnarray}
F^{BB}_{\Phi}(R)= \frac{\sin(\Phi/2)}{4\pi}\oint_{C_1} \frac{(z^2-1)z^{n-1}}{f_{\Phi}(z)} dz,
\end{eqnarray}
where $f_{\Phi}(z) ={\cos(\Phi/2)z^2+2z+\cos(\Phi/2)}$ and $C_1$ is the unit circle, $R=na$ and $n\ge 1$. 
To calculate $F^{BB}_{\Phi}(R)$, we perform a standard residue calculation for simple poles.

The flux dependent poles are $z_{\pm}=\frac{1}{\cos(\Phi/2)}\left[-1\pm \sin(\Phi/2)\right]$ where, for $ 0\le \Phi \le \pi$, 
$z^{+}$ (resp. $z^-$) is inside (resp. outside) the contour.\\
Thus we find, 
\begin{eqnarray}
I_{BB}^{00}(R)=-\frac{|JS|}{8|z_{+}|^2} \frac{\sin^2(\Phi/2)}{\left[1+\sin(\Phi/2)\right]^2}e^{-\frac{R\ln{|z_+|^{-2}}}{a}}.
\end{eqnarray} 
Thus, the decaying length $\xi(\Phi)$ is given by,
\begin{eqnarray}
\xi(\Phi)=\frac{a}{2|\ln{|z_+|}|}.
\end{eqnarray}
\section{$J_{BB}(R)$ for $\Phi=\pi$ as a function of JS}
\label{AppendixB}
\renewcommand\thefigure{B.\arabic{figure}}    
\setcounter{figure}{0} 
\begin{figure}[ht]
\centering
\includegraphics[clip, width=0.95\columnwidth]{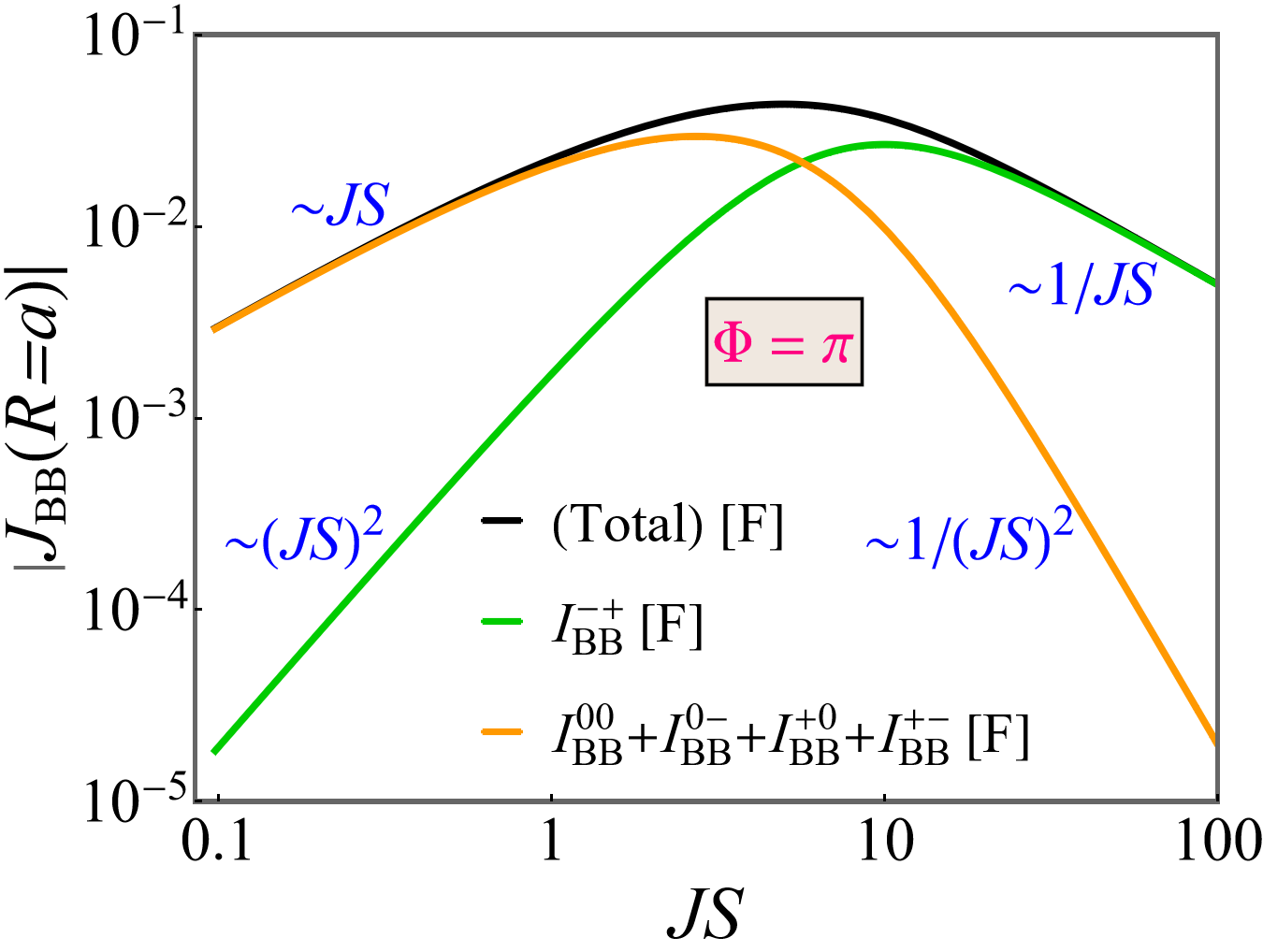}
\caption{$\vert J_{BB}(R=a)\vert$ and its contributions as a function of 
$JS$ for $\Phi=\pi$. $\vert J_{BB}(R=a)\vert$ corresponds to the black line, 
$|I^{-+}_{BB}|$ is plotted in green and the sum of the other contributions 
$|I^{00}_{BB}+I^{0-}_{BB}+I^{+0}_{BB}+I^{+-}_{BB}|$ appears in orange. 
`[F]' means that the coupling is ferromagnetic. The scaling in the weak 
and strong coupling regime is depicted as well.}
\label{JBB_vs_JS_scaling}
\end{figure}
As discussed in the main text of the manuscript, for the special case of $\Phi=\pi$, all the bands are dispersionless which 
allows for the analytical calculation of the couplings and its different contributions. 
Note that in this appendix, we use the same notations as those defined in the main text and in Appendix~\ref{AppendixA}.

In Fig.~\ref{JBB_vs_JS_scaling}, we have depicted $|J_{BB}(R=a)|$ as a function of $JS$. As discussed in the main text, the coupling 
$(B,B)$ is dominated by the FB-FB contribution ($I_{BB}^{00}$) when $JS \le 1$, while it is controlled by the single contribution 
$I_{BB}^{-+}$ when $JS \gg 1$. We now propose to give the analytical expression of $I_{BB}^{00}$ and $I_{BB}^{-+}$.
Using the expression of the eigenvalues and eigenvectors as given in the Appendix~\ref{AppendixA}, one finds, 
\begin{equation}
I_{BB}^{00}(R)=-\frac{|JS|}{32}\delta_{|R|,a}.
\end{equation}
\\
Similarly, we find, 
\begin{eqnarray}
I_{BB}^{-+}(R)=-\frac{(JS)^2}{8(\Delta-\frac{JS}{2})\big[1+(\frac{\Delta}{2t}+\frac{JS}{4t})^2\big]^2} \delta_{|R|,a}.
\label{eqImp}
\end{eqnarray} 
where we have introduced $\Delta=\sqrt{\frac{J^2S^2}{4}+16t^2}$. \\
These two expressions indicate that the couplings reduce to nearest-neighbor only, which could have been anticipated. 
It is as well the case for the other missing contributions. 

In the limit of large values of $JS$, Eq.~\eqref{eqImp} leads to, 
\begin{equation}
I_{BB}^{-+}(R) = -\frac{t^2}{8|JS|}\delta_{|R|,a},
\end{equation} 
which explains the scaling of $J_{BB}(R=a)$ in the large coupling regime as depicted in Fig.~\ref{JBB_vs_JS_scaling}. 


\end{document}